\documentclass[12pt]{article}
\usepackage{amsmath, amsthm, amsfonts}
\usepackage{lscape}
\usepackage{graphicx}
\usepackage{subfigure}

\def\a{\alpha}
\def\b{\beta}
\def\g{\gamma}
\def\d{\delta}

\def\m{\mu}
\def\n{\nu}

\def\L{\Lambda}

\def\be{\begin{equation}}
\def\ee{\nonumber\end{equation}}
\newcommand{\eel}[1]{\label{#1}\end{equation}}
\def\bea{\begin{eqnarray}}
\def\eea{\nonumber\end{eqnarray}}
\newcommand{\eeal}[1]{\label{#1}\end{eqnarray}}
\def\nn{\nonumber\\}

\def\={&=&}
\def\+{&&+}
\def\-{&&-}
\def\ba{\begin{array}}
\def\ea{\end{array}}

\begin{document}

\title{{\LARGE New families of integrable two-dimensional systems with
quartic second integrals} \vspace{0.2in}}
\author{H M Yehia$^{a,}$\thanks{%
E-mail: hyehia@mans.edu.eg} and A M Hussein$^{b,}$\thanks{%
E-mail: ahussein@sci.kfs.edu.eg} \vspace{0.2in} \\
%EndAName
\ $^{a}$ Department of Mathematics, Faculty of Science, \\
Mansoura University, Mansoura 35516, Egypt\\
\vspace{0.1in} $^{b}$ Department of Mathematics, Faculty of science, \\
Kafrelsheikh University, Kafr El-Sheikh 33516, Egypt,\\
and\\
Faculty of Science and Arts, King Khaled University, Al-Namas\\
61977, P.O. Box 101, Saudi Arabia\thanks{%
Present address}\\
}
\maketitle

\begin{abstract}
\noindent The method introduced in (Yehia H M 2006 J. Phys. A: Math. Gen.%
\textbf{\ 39} 5807-5824) and (Yehia H M 2012 J. Phys. A: Math. Gen.\textbf{\
45} 395209) is extended to construct new families of several-parameter
integrable systems, which admit a complementary integral quartic in the
velocities. A total of 12 new systems are obtained, with a number of
parameters ranging from 7 up to 16 parameters.
\end{abstract}

\section{Introduction}

Over the last few decades, enormous efforts have been dedicated to answer a
fundamental question; to determine whether a mechanical system is integrable
and how can one find integrals of the motion, if they exist? In fact, there
is no systematic method for doing that, even for integrals of the simplest
functional form, polynomial in the velocities and for the simplest
configuration space, the 2D Euclidean plane. The only possible way is to
compare one's system with available tables of known integrable cases in
different areas of interest. A fairly complete review of methods and the
small list of known integrable potentials in the Euclidean plane with an
integral polynomial in the velocities up to 1986 can be found in
Hietarinta's article \cite{Hiet:1987}. The list of cases added after that
date is even smaller. Just few more cases of systems in the plane were
obtained in few works (see e.g. \cite{Sen:1987a}, \cite{Sen:1987b} and \cite%
{Karl:2000,Karl:2002}).

The matter becomes much harder for integrable systems whose configuration
space is more general, e.g. Riemannian, 2D manifols. For a long time the
list of those cases consisted of the separable (Liouville) systems and the
few known cases of rigid body dynamics.

The method introduced by Yehia in \cite{Yehia:1986} has been most successful
in constructing new families of integrable two-dimensional mechanical
systems with second integrals polynomial in velocities with degree ranging
up to six: quadratic \cite{Yehia:1992,Yehia:2007b}, cubic \cite{Yehia:1986}, 
\cite{Yehia:2002}, quartic \cite{Yehia:2006a,Yehia:2006b}. Most known cases
with a quartic integral were recovered as special cases corresponding to
certain choices of the parameters from the so-called \emph{master} system
involving 21 arbitrary parameters. Another system with 16 free parameters
was obtained in \cite{Yehia:2012}. The results of \cite{Yehia:2006b} and 
\cite{Yehia:2012} have not only restored the famous, Kowalevski's integrable
case of rigid body dynamics \cite{Kowal:1889} and the case due to Chaplygin
of motion of a body in a liquid \cite{Chapl:1903}, but also introduced
several new integrable cases that generalized those two cases by adding
certain terms to the potential in each case \cite{Yehia:2006a} - \cite%
{Yehia:2012} and \cite{YM13}.

Yehia's method consists in two steps. The first is constructing the basic
system integrable on its zero-energy level and the second is the
interpretation of the energy constant and the standard time variable. This
usually gives the freedom to introduce several additional parameters to the
structure of the system. More details on this can be found in \cite%
{Yehia:2006a,Yehia:2012,Yehia:2013}.

The present paper is devoted to construction of integrable systems which
admit an integral quartic in velocities. It is a continuation of \cite%
{Yehia:2006a} and \cite{Yehia:2012}. Systematic application of an extension
of the method of the last papers resulted in the construction of 14 systems
with a quartic invariant, of which 12 systems are new. The new systems
involve several parameters, ranging in number up to 11 parameters. Those
systems are here classified.

\subsection{Formulation of the problem}

Consider the natural conservative mechanical system described by the
Lagrangian 
\begin{eqnarray}
L=\frac{1}{2}\sum_{i,j=1}^2a_{ij}\dot{q}_i\dot{q}_j-V,  \label{L-0}
\end{eqnarray}
where $a_{ij},\;V$ are certain functions of the generalized coordinates $%
q_1,\;q_2$ only. Clearly, the system (\ref{L-0}) admits the energy integral 
\begin{eqnarray}
H=\frac{1}{2}\sum_{i,j=1}^2a_{ij}\dot{q}_i\dot{q}_j+V=h,  \label{H-0}
\end{eqnarray}
where $h$ denotes the arbitrary energy parameter. The most general form of a
quartic integral of (\ref{L-0}) is 
\begin{eqnarray}
I=\sum_{i=0}^{4}C_{4,i}\dot{q}_1^i\dot{q}_2^{4-i} +\sum_{i=0}^{2}C_{2,i}\dot{%
q}_1^i\dot{q}_2^{2-i} +C_0  \label{I-0}
\end{eqnarray}
where $C_{ij},\;C_0$ are functions in $q_1,\;q_2$.

The problem is to determine the 13 unknown functions $\{\mbox{\sl g}%
_{ij}\},~V,~\{C_{4,i}\},~\{C_{2,i}\},~C_0$ such that $dI/dt=0$ in virtue of
the equations of motion derived from the Lagrangian (\ref{L-0}).

As was shown in \cite{Yehia:1986} and recently in \cite{Yehia:2012},
whenever a natural 2D mechanical system admits an integral of motion quartic
in velocities, this system can always be reduced in certain isometric
coordinates $\xi,\eta$ and a time parametrization $\tau$ to a ficticious
system described by the Lagrangian 
\begin{eqnarray}
L=\frac{1}{2}\left(\xi^{\prime 2}+\eta^{\prime 2}\right)+U,\quad
U=\Lambda(h-V),  \label{L-1}
\end{eqnarray}
restricted to its zero-energy level 
\begin{eqnarray}
\xi^{\prime 2}+\eta^{\prime 2}+2U=0,  \label{H-1}
\end{eqnarray}
where the prime denotes differentiation with respect to the parameter $\tau$%
, $h$ and $V$ are the energy constant and the potential function of the
original system, and $\Lambda(\xi,\eta)$ is a conformal factor which depends
on the metric of the configuration space. The quartic integral is
simultaneously written in the following simple form involving only three
unknown functions instead of nine in (\ref{I-0}): 
\begin{eqnarray}
I=\xi^{\prime 4}+P\xi^{\prime 2}+Q\xi^\prime\eta^\prime+R=\mbox{const.}
\label{I}
\end{eqnarray}
All the functions involved are expressed in terms of an auxiliary function $%
F(\xi,\eta)$, which is a solution of the nonlinear equation 
\begin{eqnarray}
\frac{\partial^2F}{\partial\xi\partial\eta} \left(\frac{\partial^4F}{%
\partial\xi^4}-\frac{\partial^4F}{\partial\eta^4}\right) +3\left(\frac{%
\partial^3F}{\partial\xi^3}\frac{\partial^3F}{\partial\xi^2\partial\eta} -%
\frac{\partial^3F}{\partial\eta^3}\frac{\partial^3F}{\partial\eta^2\partial%
\xi}\right)  \notag \\
+2\left(\frac{\partial^2F}{\partial\xi^2}\frac{\partial^4F}{%
\partial\xi^3\partial\eta} -\frac{\partial^2F}{\partial\eta^2}\frac{%
\partial^4F}{\partial\eta^3\partial\xi}\right)=0,  \label{PDE}
\end{eqnarray}
which called the \textit{resolving equation}. In terms of $F$, three of the
unknown functions of the problem, namely $P,\;Q$ and $U$, are expressed as 
\begin{eqnarray}
P=\frac{\partial^2 F}{\partial\xi^2},\quad Q=-\frac{\partial^2F}{%
\partial\xi\partial\eta},\quad U=-\frac{1}{4}\left(\frac{\partial^2F}{%
\partial\xi^2}+\frac{\partial^2F}{\partial\eta^2}\right),  \label{F2U}
\end{eqnarray}
while the function $R$ is given, up to an additive constant, by the
quadrature 
\begin{eqnarray}
R(\xi,\eta) =-\int Q\frac{\partial U}{\partial \xi} d\eta -\int\left[2P 
\frac{\partial U}{\partial \xi} +Q\frac{\partial U}{\partial \eta}+2U \frac{%
\partial Q}{\partial \eta}\right]_0d\xi,  \label{R}
\end{eqnarray}
where $[]_0$ means that the expression in the bracket is computed for $\eta$
taking an arbitrary constant value $\eta_0$ (say).

The set of solutions of (\ref{PDE}) generates all systems of the type (\ref%
{L-1}) having an integral of the form (\ref{I}) on the zero level of their
energy integral. Affecting all possible conformal mappings of the complex $%
\zeta=\xi+i\eta$ plane followed by a general point transformation to the
generalized coordinates $q_1, q_2$ with a suitable change of the time
variable we obtain all systems of the general form on two-dimensional
Riemannian (or pseudo-Riemannian) manifolds, having a quartic integral on
the zero level of their energy integral, i.e. \textit{conditional systems}.

The original system can now be expressed in terms of the coordinates $%
\xi,\eta$ and the natural time $t$ by the Lagrangian 
\begin{eqnarray}
L^*=\frac{1}{2}\Lambda\left(\dot{\xi}^2+\dot{\eta}^2\right)-V.
\end{eqnarray}
The quartic integral now takes the form 
\begin{eqnarray}
I^*=\Lambda^4\dot{\xi}^4+\Lambda^2\left(P\dot{\xi}^2+Q\dot{\xi}\dot{\eta}%
\right)+R=\mbox{const.}  \label{I*}
\end{eqnarray}

\subsection{The choice of $\Lambda$}

\label{subsec:uncond-sys} To construct systems that integrable on all energy
levels, the functions $U$ obtained from (\ref{F2U}) must have a structure in
which the energy constant $h$ of the original system enters linearly as a
parameter. Any parameter that appears only as linear multiplier in a certain
term of the potential can be identified as the energy parameter $h$ and its
cofactor as the function $\Lambda$, and we can proceed through an inverse
time transformation to construct a set of general integrable systems valid
on arbitrary energy level, i.e. unconditional systems. The general situation
however assumes $U$ to have a set of linear multipliers $h_i$. Then, the
Lagrangian can be written as (see e.g. \cite{Yehia:2012, Yehia:2013}) 
\begin{eqnarray}
L=\frac{1}{2}\left(\xi^{\prime 2}+\eta^{\prime
2}\right)-\sum_{i=1}^nh_iU_i(\xi,\eta),  \label{L_Ui}
\end{eqnarray}
which admits the quartic integral (\ref{I}) on the zero level of the energy
integral 
\begin{eqnarray}
H=\frac{1}{2}\left(\xi^{\prime 2}+\eta^{\prime
2}\right)+\sum_{i=1}^nh_iU_i(\xi,\eta)=0.  \label{H_Ui-0}
\end{eqnarray}
Introducing new arbitrary parameters $\alpha_i,\beta_i$ into (\ref{L_Ui}) by
the substitution $h_i=\alpha_i-\beta_ih$, we get 
\begin{eqnarray}
L=\frac{1}{2}\left(\xi^{\prime 2}+\eta^{\prime 2}\right)
-\left(\sum_{i=1}^n\beta_iU_i\right) \left(\frac{\sum_{i=1}^n\alpha_iU_i}{%
\sum_{i=1}^n\beta_iU_i}-h\right).  \label{L_Ui2}
\end{eqnarray}
Making the change of the independent variable $\tau$ to the original time $t$
according to the relation 
\begin{eqnarray}
t=\int \sum_{i=1}^n\beta_i U_i\;d\tau,
\end{eqnarray}
we obtain the Lagrangin $L_1=L^*+h$, where 
\begin{eqnarray}
L^*=\frac{1}{2}\left(\sum_{i=1}^n\beta_iU_i\right)\left(\dot{\xi}^2+\dot{\eta%
}^2\right) -\frac{\sum_{i=1}^n\alpha_iU_i}{\sum_{i=1}^n\beta_iU_i},
\label{L*}
\end{eqnarray}
while the energy integral (\ref{H_Ui-0}) is transformed to 
\begin{eqnarray}
\frac{1}{2}\left(\sum_{i=1}^n\beta_iU_i\right)\left(\dot{\xi}^2+\dot{\eta}%
^2\right) +\frac{\sum_{i=1}^n\alpha_iU_i}{\sum_{i=1}^n\beta_iU_i}=h.
\label{H_Ui}
\end{eqnarray}
The second integral of $L_1$ is obtained from (\ref{I}).

Now, discarding the free additive parameter $h$ from $L_1$ reduces it to $%
L^* $. Since the zero level of energy integral of $L_1$ is the $h$-level for 
$L^* $, as determined by (\ref{H_Ui}), the Lagrangian $L^*$ admits the
second integral (\ref{I*}) on its $h$-level of energy. Finally, one can use
the energy integral (\ref{H_Ui}) to eliminate $h$ from (\ref{I*}) and then
get a form of the second integral free of the energy parameter.

\section{New solutions of the resolving equation}

The discussion in \cite{Yehia:2006b} (see also \cite{Yehia:2012}) showed
that, in certain circumstances, the original isometric variables $\xi,\;\eta$
are not practically suitable for solving the equation, and that the
symmetric separation solution discussed in \cite{Yehia:2006b} can be more
conveniently expressed in the coordinates $p$ and $q$ defined by 
\begin{eqnarray}
\xi=\int^{p}\frac{dz}{\sqrt[4]{a_4z^4+a_3z^3+a_2z^2+a_1z+a_0}},  \notag \\
\eta=\int^{q}\frac{dz}{\sqrt[4]{a_4z^4+b_3z^3+b_2z^2+b_1z+b_0}},  \label{x&h}
\end{eqnarray}
where $a_4,a_3,a_2,a_1,a_0,b_3,b_2,b_1,b_0$ are arbitrary constants. The
integrable system constructed in \cite{Yehia:2006b}, the master system,
represents the solution of (\ref{PDE}) when 
\begin{eqnarray}
F(\xi,\eta)\=\int\!\!\!\!\int\!f(\xi)~d\xi d\xi +\int\!\!\!\!\int\!
g(\eta)~d\eta d\eta+\nu pq,  \label{F4Master}
\end{eqnarray}
where 
\begin{eqnarray}
f(\xi)=\frac{\frac{1}{4}b_3p^3+4Ap^2+4C_1p+4C_0}{\sqrt{%
a_4p^4+a_3p^3+a_2p^2+a_1p+a_0}},  \notag \\
g(\eta)=\frac{\frac{1}{4}a_3q^3+4Aq^2+4D_1q+4D_0}{\sqrt{%
a_4q^4+b_3q^3+b_2q^2+b_1q+b_0}},  \label{f&g}
\end{eqnarray}
where $\nu,A,C_1,C_0,D_1,D_0$ are arbitrary parameters. Another solution of
the resolving equation is obtained in \cite{Yehia:2012} assuming $F$ in the
form 
\begin{eqnarray}
F(\xi,\eta)\=\int\!\!\!\!\int\!f(\xi)~d\xi d\xi+\int\!\!\!\!\int\!
g(\eta)~d\eta d\eta +\nu pq+\nu_1p^2q^2,  \label{F4NewMaster}
\end{eqnarray}
where $\nu$ and $\nu_1$ are arbitrary constants. The above two choices led
to the construction of two systems integrable on all energy levels and
involving a total number of parameters 21 and 16 respectively. Special cases
of the two system admit interpretation in particle and rigid body dynamics
(See \cite{Yehia:2006b,Yehia:2012} for detail).

The main object of the present work is to extend further the method of \cite%
{Yehia:2006b,Yehia:2012} to construct and classify integrable systems
corresponding to the generalized ansatz 
\begin{eqnarray}
F(\xi,\eta)\=\int\!\!\!\!\int\!f(\xi)~d\xi d\xi+\int\!\!\!\!\int\!
g(\eta)~d\eta d\eta +\sum_{i=2}^4\sum_{j=2}^i\nu_{j,i-j}p^{i}q^{i-j},
\label{F-O(4)}
\end{eqnarray}
possibly with certain restrictions on the parameters involved in $%
f(\xi),g(\eta)$ as given in (\ref{f&g}).

Substituting (\ref{F-O(4)}) into equation (\ref{PDE}), and making use of (%
\ref{x&h}), we get a polynomial expression of the sixth degree in $p,q$ that
must vanish. This yields a system of 27 polynomial equations in the 26
parameters of the problem; $\{A,C_{0},C_{1},D_{0},D_{1},a_{i}(i=0,\dots
,4),b_{j}(j=0,\dots ,3),\nu _{ij}(2\leq i\leq j\leq 4)\}$. The system of
polynomial equations is solved using the MAPLE computer algebra package and
we obtained 59 distinct solutions, i.e 59 working combinations of the
parameters that may lead to the construction of integrable systems with a
quartic second integral. It turned out that for 17 solutions the
corresponding integrable systems are separable and then they admit integrals
quadratic in velocities and will not be considered further. Moreover, due to
the symmetric way in which groups of parameters are associated to the
variables p, q, there exist 16 symmetry relations between the remaining 42
solutions. This reduced the number of independent solutions to 26. Finally,
it turned out that 13 cases can be obtained by assuming special values of
parameters in the other 13. Thus, the final number of different systems is
thirteen, the number we are going now to classify and put in a form as
simple as possible.

\section{The basic integrable systems}

In this section we tabulate the 13 basic integrable systems with a quartic
integral. Those are the conditional ones, valid on their zero-energy levels.
We first note that

\begin{enumerate}
\item The first 5 systems could be expressed explicity in terms of the
cartesian coordinates $\xi,\eta$ in a Euclidean plane. This possibility was
ignored, since the resulting expressions contained rational powers which
makes the potential and the complementary integral more complicated.

\item The system No. 3 (given by the Lagrangian \ref{SYSI.03}) already
involves an additional arbitrary parameter $d$, which can be interpreted as
an energy parameter. This system is thus unconditionally integrable, but
still we can add more parameters to its structure in the next section.

\item The two systems by numbers 11 and 13 were given earlier in \cite%
{Yehia:2012} and \cite{Yehia:2006b} respectively. They are given here only
for completeness of the results.
\end{enumerate}

For each case in table I we give the Lagrangian and the complementary
integral valid on its zero-energy level. The systems are classified in Table
I according to the number of arbitrary parameters entering into their
structure.

\newpage 
\begin{landscape}
\noindent
\textbf{Table I.} 
\textit{Basic (conditional) systems}  
\vspace*{-.3cm}\\
\rule[-0.1cm]{21.5cm}{0.01cm}\\[-.4cm]
\rule[-0.1cm]{21.5cm}{0.03cm}\\ 
{\small
\begin{eqnarray}
 1.\quad % System No. 1 (L)
 &L=&\frac{1}{2}\left(p^4p^{\prime 2}+q^4q^{\prime 2}\right)
     -(p^2+q^2)\left\{
       a\left[\left(p^2-q^2\right)^4\left(5\left(\frac{p}{q}-\frac{q}{p}\right)^2
              +\frac{\d}{p^2q^2}\right)
       +\left(12p^4-16p^2q^2+12q^4-\d\right)^2\right]\right.\nn
 &  &\left.+b\left(\frac{p}{q}-\frac{q}{p}\right)^2
     +\frac{c}{p^2q^2}\right\},\nn
 &I=&\left\{\frac{1}{2}~p^4p^{\prime 2}+119ap^{10}+27a(5q^4-\d)p^6
            -\left[a\left(195q^8-10\d q^4-\d^2\right)+b\right]p^2
      +\frac{aq^8(5q^4+\d)+bq^4+c}{p^2}\right\}^2\nn
 &  &-4p^3q^3\left\{a\left[15p^8-2(39q^4-\d)p^4+15q^8+2q^4\d\right]+b\right\}p^\prime q^\prime
     +2964a^2p^{20}-1800q^2a^2p^{18}
     -4a^2(12495q^4-94\d)p^{16}\nn
 &  &+480a^2q^2(39q^4-\d)p^{14}
     +4a\left[a\left(17130q^8+1240\d q^4-13\d^2\right)+122b\right]p^{12}
     -16aq^2\left[a(3267q^8-126\d q^4+2\d^2)+15b\right]p^{10}\nn
 &  &+4a\left[17130aq^{12}-5484a\d q^8+(141a\d^2-282b)q^4-a\d^3-6b\d+78c\right]p^8
     +32aq^2(15aq^8+2a\d q^4+b)(39q^4-\d)p^6\nn
 &  & -\left\{a(49980aq^{16}-4960a\d q^{12}-564(a\d^2-2b)q^8+24(a\d^3-10b\d+10c)q^4+4\d(b\d+4c))
       -4b^2\right\}p^4\nn
 &  &-8q^2(15aq^8+2a\d q^4+b)^2p^2
     +4q^4\left\{
        a\left[741aq^{16}+94a\d q^{12}-(13a\d^2-122b)q^8-(a\d^3+6b\d-78c)q^4-b\d^2-4c\d\right]
        +b^2\right\}\nn
\label{SYSI.01}
\end{eqnarray}
%%%
\begin{eqnarray}
 2.\quad % System No. 2 (L)
 &L=&\frac{1}{2}\left(pp^{\prime 2}+qq^{\prime 2}\right)
        -\left(\frac{1}{q}+\frac{1}{p}\right)
         \left\{(p+q)^4\left[a\left(5p^2+6pq+5q^2\right)+b\right]
                +c(p+q)^2+d\right\},\nonumber\\
 &I=&\left\{\frac{1}{2}~pp^{\prime 2}+31ap^5+5(27aq^2+b)p^3+(85aq^4+10bq^2+3c)p
            +\frac{5aq^6+bq^4+cq^2+d}{p}\right\}^2\nn
   &&-4pq\left[a\left(5p^2+3q^2\right)\left(3p^2+5q^2\right)+2\left(p^2+q^2\right)b+c\right]
       p^\prime q^\prime
     -1292a^2p^{10}-1800a^2qp^9-12a(1085aq^2+34b)p^8\nn
   &&-480aq(17aq^2+b)p^7
     -8\left[a(4355aq^4+380bq^2+33c)+4b^2\right]p^6
     -16q\left[a(803aq^4+98bq^2+15c)+2b^2\right]p^5\nn
   &&-8\left[4355a^2q^6+674abq^4+(159ac+20b^2)q^2+17ad+5bc\right]p^4
     -32q\left[255a^2q^6+49abq^4+(17ac+2b^2)q^2+bc\right]p^3\nn
   &&-4\left[3255a^2q^8+760abq^6+2(159ac+20b^2)q^4+4(15ad+7bc)q^2+4bd+3c^2\right]p^2\nn
   &&-8q\left[225a^2q^8+60abq^6+2(15ac+2b^2)q^4+4bcq^2+c^2\right]p\nn
   &&-4q^2\left[323a^2q^8+102abq^6+2(33ac+4b^2)q^4+2(17ad+5bc)q^2+4bd+3c^2\right].
\label{SYSI.02}
\end{eqnarray}
%%%
\newpage
\begin{eqnarray}
 3.\quad % System No. 3 (L)
 &L=&\frac{1}{2}\left(p^4p^{\prime 2}+q^4q^{\prime 2}\right)
     -a\left[\frac{9p^6+2q^6}{2p^2}+30\d p^2q^3+\d^2\left(9p^6+64q^6\right)\right]
     -\frac{b}{p^2}
     -c\left(16\d q^3+3p^2\right)-d,\nn
 &I=&\left[\frac{1}{2}~p^4p^{\prime 2}+9a\d^2p^6+3(10a\d q^3+c)p^2+\frac{aq^6+b}{p^2}\right]^2
     -6ap^3q^2(3\d p^4+q^3)p^\prime q^\prime
     -162a^2\d^2p^{10}\nn
   &&-\frac{27a}{4}(128a\d^3q^3+16c\d^2+3a)p^8
     -108a^2\d q^3p^6-9a(128a\d^2q^6+24c\d q^3+d)p^4
     -18a^2q^6p^2
     -12aq^3\left(8a\d q^6+cq^3+4b\d\right).\nn
\label{SYSI.03}
\end{eqnarray}
%%%
\begin{eqnarray}
4.\quad % System No. 4 (L)
 &L=&\frac{1}{2}\left(pp^{\prime 2}+qq^{\prime 2}\right)
     -a(p+q)^3\left[\frac{(p+q)^4}{pq}+4(p-q)^2\right]
     -\frac{b(p+q)^5}{pq}
     -c\left(\frac{p^2}{q}+\frac{q^2}{p}\right)
     -d\left(\frac{1}{q}+\frac{1}{p}\right)
     -e(p+q),\nonumber\\
 &I=&\left\{\frac{1}{2}~pp^{\prime 2}
            +11ap^5+(27aq^2+5b)p^3+(25aq^4+10bq^2+e)p+\frac{aq^6+bq^4+cq^2+d}{p}\right\}^2\nn
   &&-4pq\left[3ap^4+2(5aq^2+b)p^2+3aq^4+2bq^2+c\right]p^\prime q^\prime
     -76a^2p^{10}
     -72a^2qp^9
     -4a(231aq^2+26b)p^8
     -96aq(5aq^2+b)p^7\nn
   &&-4\left[a(518aq^4+200bq^2+15c+e)+8b^2\right]p^6
     -16q\left[a(59aq^4+26bq^2+3c)+2b^2\right]p^5\nn
   &&-4\left[518a^2q^6+316abq^4+(33ac+15ae+40b^2)q^2+10ad+7bc+be\right]p^4
     -32q\left[15a^2q^6+13abq^4+(5ac+2b^2)q^2+bc\right]p^3\nn
   &&-4\left[231a^2q^8+200abq^6+(33ac+15ae+40b^2)q^4+(12ad+10bc+6be)q^2+4bd+ce\right]p^2\nn
   &&-8q\left[9a^2q^8+12abq^6+2(3ac+2b^2)q^4+4bcq^2+c^2\right]p\nn
   &&-4q^2\left[19a^2q^8+26abq^6+(15ac+ae+8b^2)q^4+(10ad+7bc+be)q^2+4bd+ce\right].
\label{SYSI.04}
\end{eqnarray}
%%%
\begin{eqnarray}
5.\quad % System No. 5 (L)
 &L=&\frac{1}{2}\left(p^4p^{\prime 2}+q^4q^{\prime 2}\right)
     -\left(\d p^2+q^2\right)
       \left\{a\left[4(\d p^2+q^2)^2+\left(\frac{p^3}{q}-\frac{\d q^3}{p}\right)^2\right]
                +b\left(\frac{\d q^2}{p^2}+\frac{p^2}{q^2}\right)+c\right\}
     -\frac{d}{p^2}-\frac{e}{q^2},\nn
 &I=&\left\{\frac{1}{2}~p^4p^{\prime 2}
            +a(4\d^3+1)p^6+(10a\d q^4+c\d)p^2+\frac{\d q^4(a\d q^4+b)+d}{p^2}\right\}^2
     -4\d p^3q^3\left(2a\d q^4+2ap^4+b\right)p^\prime q^\prime\nn
   &&-32a^2\d^3p^{12}
     -32a^2\d^2q^2p^{10}
     -4a\d\left[8aq^4(4\d^3+1)+8b\d^2+c\right]p^8
     -32a\d^2q^2(2a\d q^4+b)p^6\nn
   &&-4\d\left[8a^2\d q^8(\d^3+4)+2aq^4(4b\d^3+3c\d+4b)+b^2\d^2+4ae\d+bc\right]p^4\nn
   &&-8\d^2q^2\left(2a\d q^4+b\right)^2p^2
      -4\d q^4\left\{a\left[8a\d^2q^8+\d(c\d+8b)q^4+4d\right]+bc\d+b^2\right\}.
\label{SYSI.05}
\end{eqnarray}
%%%
\newpage
\begin{eqnarray}
6.\quad % System No.06 (L83)
    &L=&\frac{1}{2}\left[\frac{p^{\prime 2}}{\sqrt{a_2p^2+a_0}}
                         +\frac{q^{\prime 2}}{\sqrt{a_2q^2+a_0}}\right]\nn
    &  &-\frac{\m a_2q^3p+q^2\left[5\m (3a_2p^2+2a_0)+\m_1a_2p\right]
               +q\left[15\m a_2p^3+6\m_1a_2p^2+(2\m a_0+A)p+4\m_1a_0\right]
               +\m a_2p^4+\m_1a_2p^3+Ap^2+C_1p+C_0}
              {\sqrt{a_2p^2+a_0}}\nn
    &  &-\frac{\m a_2p^3q+p^2\left[5\m (3a_2q^2+2a_0)+\m_1a_2q\right]
               +p\left[15\m a_2q^3+6\m_1a_2q^2+(2\m a_0+A)q+4\m_1a_0\right]
               +\m a_2q^4+\m_1a_2q^3+Aq^2+C_1q+C_0}
              {\sqrt{a_2q^2+a_0}},\nn
 &I=&\frac{a_2^2}{a_2p^2+a_0}\left\{p^{\prime 2}/2+\m a_2q^3p+q^2\left[5\m (3a_2p^2+2a_0)+\m_1a_2p\right]
                +q\left[15\m a_2p^3+6\m_1a_2p^2+(2\m a_0+A)p+4\m_1a_0\right]
                +\m a_2p^4\right.\nn
 &  &\left.+\m_1a_2p^3+Ap^2+C_1p+C_0\right\}^2
    -2a_2\left[3\m a_2q^2+2a_2q(5\m p+\m_1)+3\m a_2p^2+2\m_1a_2p-10\m a_0+A\right]p^\prime q^\prime\nn
 &  &-2\left[3\m a_2q^2+2a_2q(5\m p+\m_1)+3\m a_2p^2+2\m_1 a_2p-10\m a_0+A\right]^2
         \sqrt{a_2p^2+a_0}\sqrt{a_2q^2+a_0}
-\m^2a_2^3q^6
        -2\m a_2^3(24\m p+\m_1)q^5\nn
 &  &-a_2^2\left[\m(375\m a_2p^2+66\m_1a_2p+98\m a_0+2A)+\m_1^2a_2\right]q^4
        -2a_2^2\left[344\m^2a_2p^3+158\m\m_1a_2p^2+2(5\m_1^2a_2+11\m A-20\m^2a_0)p\right.\nn
 &  &+\left.\m(C_1+38\m_1a_0)+\m_1A\right]q^3
        -a_2\left\{375\m^2a_2^2p^4+316\m\m_1a_2^2p^3
        +2a_2\left[\m(50A-210\m a_0)+27\m_1^2a_2\right]p^2
        +2a_2\left[15\m(C_1-2\m_1a_0)+11\m_1A\right]p\right.\nn
 &  &-\left.4\m(5\m a_0^2+2a_0A)+2\m_1a_2(C_1+8\m_1^2a_0)+20\m a_2C_0+A^2\right\}q^2
        -2a_2\left\{24\m^2a_2^2p^5+33\m\m_1a_2^2p^4
           +2a_2\left[\m(11A-20\m a_0)+5\m_1^2a_2\right]p^3\right.\nn
 &  &+\left.a_2\left[15\m(C_1-2\m_1a_0)+11\m A\right]p^2
        +2\left[2\m(3a_2C_0-5a_0A)+3\m_1a_2C_1+A^2\right]p+C_1(A-10\m a_0)+4\m_1a_2C_0\right\}q
        -\m^2a_2^3p^6
        -2\m\m_1a_2^3p^5\nn
 &  &-a_2^2\left[2\m(49\m a_0+A)+\m_1^2a_2\right]p^4
        -2a_2^2\left[\m(38\m_1 a_0+C_1)+\m_1 A\right]p^3
        -a_2\left\{4\m\left[5a_2C_0-a_0(5\m a_0+2A)\right]+2\m_1a_2(8\m_1a_0+C_1)+A^2\right\}p^2\nn
 &  &-2a_2\left[C_1(A-10\m a_0)+4\m_1a_2C_0\right]p
\label{SYSI.06}
\end{eqnarray}
%%%
\begin{eqnarray}
7.\quad % System No.07 (L34)
    &L=&\frac{1}{2}\left[\frac{p^{\prime 2}}{\sqrt{a_2p^2+a_0}}
                      +\frac{q^{\prime 2}}{\sqrt{a_2q^2+b_0}}\right]
        -\frac{\m a_2q^2p+q\left[2\m (3a_2p^2+2a_0)+Ap\right]+\m a_2p^3+Ap^2+C_1p+C_0}
              {\sqrt{a_2p^2+a_0}}\nn
    &  &-\frac{\m a_2p^2q+p\left[2\m (3a_2q^2+2b_0)+Aq\right]+\m a_2q^3+Aq^2+C_1q+D_0}
              {\sqrt{a_2q^2+b_0}},\nn
    &I=&\frac{a_2^2\left\{p^{\prime 2}/2+\m a_2q^2p+q\left[2\m (3a_2p^2+2a_0)+Ap\right]
                      +\m a_2p^3+Ap^2+C_1p+C_0\right\}^2}
             {a_2p^2+a_0}
         -2a_2[2\m a_2(p+q)+A]p^\prime q^\prime\nn
    &  & -2[2\m a_2(p+q)+A]^2\sqrt{a_2p^2+a_0}\sqrt{a_2q^2+b_0}
         -\m^2 a_2^3q^4
         -2\m a_2^2(10\m a_2p+A)q^3
         -a_2\left[2\m a_2\left(27\m a_2p^2+11Ap+8\m a_0+C_1\right)+A^2\right]q^2\nn
    &  & -2a_2\left[10\m^2 a_2^2p^3+11\m a_2Ap^2+2(A^2+3\m a_2C1)p+AC_1+4\m C_0\right]q
         -\m^2 a_2^3p^4-2\m a_2^2Ap^3
         -a_2\left[2\m a_2\left(8\m b_0+C_1\right)+A^2\right]p^2\nn
    &  & -2a_2(C_1A+4\m a_2D_0)p.
\label{SYSI.07}
\end{eqnarray}
\newpage
\begin{eqnarray}
8.\quad % System No. 08 (L03)
    &L=&\frac{1}{2}\left[\frac{p^{\prime 2}}{\sqrt{a_2p^2+a_1p+a_0}}
                         +\frac{q^{\prime 2}}{\sqrt{b_1q+b_0}}\right]
        -\frac{\m q(b_1^2q^2+3b_0b_1q+3b_0^2)\left(2a_2p+a_1\right)
              +81\m b_1^3p^3+9b_1^2Ap^2+C_1p+C_0}{\sqrt{a_2p^2+a_1p+a_0}}\nn
 &  &-\left(b_1q+b_0\right)^{3/2}\left(27\m b_1p+A\right),\nonumber\\
 &I=&\frac{\left[p^{\prime 2}/2+\m q(b_1^2q^2+3b_0b_1q+3b_0^2)\left(2a_2p+a_1\right)
       +81\m b_1^3p^3+9b_1^2Ap^2+C_1p+C_0\right]^2}{a_2p^2+a_1p+a_0}
       -12\m\left(b_1q+b_0\right)^2p^\prime q^\prime\nn
 &  &-72\m^2(b_1q+b_0)^{9/2}\sqrt{a_2p^2+a_1p+a_0}
        -4\m\left\{q(b_1^2q^2+3b_0b_1q+3b_0^2)\left[\m
            a_2q(b_1^2q^2+3b_0b_1q+3b_0^2)+C_1\right]\right.\nn
 &  &+\left.9b_1p(b_1q+b_0)^3(27\m b_1p+2A)\right\}.
\label{SYSI.08}
\end{eqnarray}
%%%
\begin{eqnarray}
9.\quad % System No. 09 (L17)
    &L=&\frac{1}{2}\left[\frac{p^{\prime 2}}{\sqrt{a_2p^2+a_1p+a_0}}+q^{\prime 2}\right]
        -\frac{\m q(q+\m_1)(2a_2p+a_1) +4Ap^2+C_1p+C_0}{\sqrt{a_2p^2+a_1p+a_0}}
        -8\m p-Aq(q+\m_1)-D_0,\nonumber\\
 &I=&\frac{\left[p^{\prime 2}/2+\m q(q+\m_1)(2a_2p+a_1)+4Ap^2+C_1p+C_0\right]^2}{a_2p^2+a_1p+a_0}
        -4\m\left(2q+\m_1\right)p^\prime q^\prime
        -8\m^2(2q+\m_1)^2\sqrt{a_2p^2+a_1p+a_0}\nn
 & & -4\m\left\{q(q+\m_1)\left[a_2\m q(q+\m_1)+8Ap+C_1\right]
                   +16\m p^2+(\m_1^2A+4D_0)p\right\}.
 \label{SYSI.09}
\end{eqnarray}
%%%
\begin{eqnarray}
 10.\quad % System No. 10 (L59)
    & L=&\frac{1}{2}\left[
         \frac{p^{\prime 2}}{\sqrt{a_2p^2+a_0}}+\frac{q^{\prime 2}}{\sqrt{b_2q^2+b_0}}\right]
         -\frac{\m q^2(3a_2p^2+2a_0)+\m_1a_2 pq+\m b_2p^4+Ap^2+C_0}{\sqrt{a_2p^2+a_0}}\nn
 &   &-\frac{\m p^2(3b_2q^2+2b_0)+\m_1b_2 qp+\m a_2q^4+Aq^2+D_0}{\sqrt{b_2q^2+b_0}},\nn
 &I=&\frac{\left[p^{\prime 2}/2
                  +\m q^2(3a_2p^2+2a_0)+\m_1a_2 pq+\m b_2p^4+Ap^2+C_0\right]^2}
                    {a_2p^2+a_0}
        -2(2\m qp+\m_1)p^\prime q^\prime
        -2(2\m qp+\m_1)^2\sqrt{a_2p^2+a_0}\sqrt{b_2q^2+b_0}\nn
 &  &-4\m^2(3a_2p^2+a_0)q^4-8a_2\m\m_1pq^3
        -\left[4\m(3b_2\m p^4+2Ap^2+C_0)+\m_1^2a2\right]q^2
        -4\m p(2\m b_2p^2+A)q-4\m^2b_0p^4\nn
 &  &-(\m_1^2b_2+4\m D_0)p^2.
\label{SYSI.10}
\end{eqnarray}
\newpage
\begin{eqnarray}
11.\quad % System No.11 (L43)
    &L=&\frac{1}{2}\left[\frac{p^{\prime 2}}{2\sqrt{a_4p^4+a_2p^2+a_0}}
                        +\frac{q^{\prime 2}}{2\sqrt{a_4q^4+b_2q^2+b_0}}\right]
        -\frac{\m q^2(4a_4p^4+3a_2p^2+2a_0)+\m_1 qp(2a_4p^2+a_2)+\m b_2p^4+Ap^2+C_0}
              {\sqrt{a_4p^4+a_2p^2+a_0}}\nn
 &  &-\frac{\m p^2(4a_4q^4+3b_2q^2+2b_0)+\m_1 pq(2a_4q^2+b_2)+\m a_2q^4+Aq^2+D_0}
            {\sqrt{a_4q^4+b_2q^2+b_0}},\nn
 &I=&\frac{\left[p^{\prime 2}/2+\m q^2(4a_4p^4+3a_2p^2+2a_0)
                    +\m_1 qp(2a_4p^2+a_2)+\m b_2p^4+Ap^2+C_0\right]^2}
             {a_4p^4+a_2p^2+a_0}
        -2(2\m pq+\m_1)p^\prime q^\prime\nn
 &  &-2(2\m pq+\m_1)^2\sqrt{a_4q^4+b_2q^2+b_0}\sqrt{a_4p^4+a_2p^2+a_0}
        -4\m^2(6a_4p^4+3a_2p^2+a_0)q^4
        -8\m_1\m p(3a_4p^2+a_2)q^3\nn
 &  &-(12\m^2b_2p^4+2(3\m_1^2a_4+4\m A)p^2+\m_1^2a_2+4\m C_0)q^2
        -4\m_1p(2\m b_2p^2+A)q
        -4\m^2b_0p^4-(\m^2b_2+4\m D_0)p^2.
 \label{SYSI.11}
\end{eqnarray}
%%%
\begin{eqnarray} 
12.\quad % System No.12 (L41)
    &L=&\frac{1}{2}\left[\frac{p^{\prime 2}}{\sqrt{a_3p^3+a_2p^2+a_1p+a_0}}
                    +\frac{q^{\prime 2}}{\sqrt{b_2q^2+b_0}}\right]
        -\frac{4\m q^2(3a_3p^2+2a_2p^2+a_1)+64\m b_2p^3+4Ap^2+C_1p+C_0}
            {\sqrt{a_3p^3+a_2p^2+a_1p+a_0}}\nn
 &  &-\frac{16\m p\left(3b_2q^2+2b_0\right)+\m a_3q^4+Aq^2+D_0}
              {\sqrt{b_2q^2+b_0}},\nn
    &I=&\frac{\left[p^{\prime 2}/2+4\m q^2(3a_3p^2+2a_2p+a_1)+64\m b_2p^3+4Ap^2+C_1p+C_0\right]^2}
             {a_3p^3+a_2p^2+a_1p+a_0}
        -32\m q p^\prime q^\prime
        -512\m^2 q^2\sqrt{a_3p^3+a_2p^2+a_1p+a_0}\sqrt{b_2q^2+b_0}\nn
 &  &-16\m\left[4\m q^4(3a_3p+a_2)+q^2(192\m b_2p^2+8Ap+C_1)+64\m b_0p^2+4D_0p\right].
\label{SYSI.12}
\end{eqnarray}
%%%
\begin{eqnarray}
  13.\quad % System No.8 (L42)
    &L=&\frac{1}{2}\left[
          \frac{\dot{p}^2}{\sqrt{a_4p^4+a_3p^3+a_2p^2+a_1p+a_0}}
        +\frac{\dot{q}^2}{\sqrt{a_4q^4+b_3q^3+b_2q^2+b_1q+b_0}}\right]\nn
 &  &-\frac{\m q\left(4a_4p^3+3a_3p^2+2a_2p+a_1\right)
              +\m b_3p^3+Ap^2+C_1p+C_0}{\sqrt{a_4p^4+a_3p^3+a_2p^2+a_1p+a_0}}
         -\frac{\m p\left(4a_4q^3+3b_3q^2+2b_2q+b_1\right)
              +\m a_3q^3+Aq^2+D_1q+D_0}{\sqrt{a_4q^4+b_3q^3+b_2q^2+b_1q+b_0}},\nn
 &I=&\frac{\left[\dot{p}^2/2
         +\m q(4a_4p^3+3a_3p^2+2a_2p+a_1)+\m b_3p^3+Ap^2+C_1p+C_0\right]^2} 
        {a_4p^4+a_3p^3+a_2p^2+a_1p+a_0}-4\m\dot{p}\dot{q}\nn
 &  &-8\m^2\sqrt{a_4p^4+a_3p^3+a_2p^2+a_1p+a_0}\sqrt{a_4q^4+b_3q^3+b_2q^2+b_1q+b_0}\nn
 &  &-4\m\left[\m q^2(3a_3p+6a_4p^2+a_2)+q(3\m b_3p^2+2Ap+C_1)+\m b_2p^2+D_1p\right].
\label{SYSI.13}
\end{eqnarray}
}\noindent
\rule[-0.1cm]{20.65cm}{0.01cm}\\[-.45cm]
\rule[-0.1cm]{20.65cm}{0.03cm}
\end{landscape}

\newpage

\section{Classification of the unconditional integrable systems}

In Table II below we list the most general deformations of the basic
integrable systems of the preceding section into their unconditional
counterparts, valid for arbitrary initial conditions. Those systems are
constructed in the way described in \S 2. For each system we give the number
of parameters in its structure, the final form of the Lagrangian $L^{\ast }$
(written simply as $L$) and the conformal factor $\Lambda $. The
complementary integral $I^{\ast }$ will not be written down. It can be
obtained for each case from the corresponding integral $I$ of the
corresponding basic system by performing three steps:

\begin{enumerate}
\item Substituting $p^{\prime }$ and $q^{\prime }$ by $\Lambda p^{\prime }$
and $\Lambda q^{\prime }$, respectively.

\item Changing the energy-like parameters in $I$ according to: 
\begin{eqnarray*}
&&\mu =\nu -\alpha h,~\mu _{1}=\gamma -\beta h,~A=h_{2}-\alpha _{2}h, \\
&&C_{1}=h_{1}-\alpha _{1}h,~C_{0}=h_{0}-\alpha _{0}h,~D_{1}=k_{1}-\beta
_{1}h,~D_{0}=k_{0}-\beta _{0}h.
\end{eqnarray*}

\item The total energy parameter $h$, appearing in $I$ after the last
substitutions, is replaced by the energy integral corresponding to the
Lagrangian $L^{\ast }$.
\end{enumerate}

\textbf{Remark}: The potential of the system number 9 in Table I involves
several parameters in a linear way, but it is a bilinear function in the
parameters $\mu _{1},A$ and thus one can use at a time either $\mu _{1}$ or $%
A$ as an energy-like parameter. Thus, this system generates the two distinct
unconditional systems occupying numbers 9 and 10 in Table II.

\newpage 
\begin{landscape}
\noindent
\textbf{Table II.} 
\textit{Unconditional generalization} 
\vspace*{-.3cm}\\
\rule[-0.1cm]{20.65cm}{0.01cm}\\[-.4cm]
\rule[-0.1cm]{20.65cm}{0.03cm}\\[.5cm] 
{\small
1 - Number of parameters: 7
\begin{eqnarray}
 &L=&\frac{1}{2}\L\left(p^4\dot{p}^2+q^4\dot{q}^2\right)
     -\frac{1}{\L}(p^2+q^2)\left\{
       h_1\left[\left(p^2-q^2\right)^4\left[5\left(\frac{p}{q}-\frac{q}{p}\right)^2
                +\frac{\d}{p^2q^2}\right]
       +\left(12p^4-16p^2q^2+12q^4-\d\right)^2\right]\right.\nn
 & &+\left.h_2\left(\frac{p}{q}-\frac{q}{p}\right)^2
     +\frac{h_3}{p^2q^2}\right\},\nonumber\\
\nonumber\\
 &\L=&(p^2+q^2)\left\{
       \a_1\left[\left(p^2-q^2\right)^4\left[5\left(\frac{p}{q}-\frac{q}{p}\right)^2
                +\frac{\d}{p^2q^2}\right]
       +\left(12p^4-16p^2q^2+12q^4-\d\right)^2\right]+\a_2\left(\frac{p}{q}-\frac{q}{p}\right)^2
     +\frac{\a_3}{p^2q^2}\right\}.
\end{eqnarray}
2 - Number of parameters: 8
\begin{eqnarray}
    &L=&\frac{1}{2}\L\left(p\dot{p}^2+q\dot{q}^2\right)
        -\frac{1}{\L}\left\{\left(\frac{1}{q}+\frac{1}{p}\right)
         \left\{(p+q)^4\left[h_1\left(5p^2+6pq+5q^2\right)+h_2\right]
                +h_3(p+q)^2+h_4\right\}\right\},\nonumber\\
 &\L=&\left(\frac{1}{q}+\frac{1}{p}\right)
         \left\{(p+q)^4\left[\a_1\left(5p^2+6pq+5q^2\right)+\a_2\right]
                +a_3(p+q)^2+a_4\right\}.
\end{eqnarray}
3 - Number of parameters: 9
\begin{eqnarray}
    &L=&\frac{1}{2}\L\left(p^4\dot{p}^2+q^4\dot{q}^2\right)
       -\frac{1}{\L}\left\{h_1\left[\frac{9p^6+2q^6}{2p^2}+30\d p^2q^3+\d^2\left(9p^6+64q^6\right)\right]
       +\frac{h_2}{p^2}
       +h_3\left(16\d q^3+3p^2\right)+h_4\right\},\nonumber\\
 &\L=&\a_1\left[\frac{9p^6+2q^6}{2p^2}+30\d p^2q^3+\d^2\left(9p^6+64q^6\right)\right]
       +\frac{\a_2}{p^2}
       +\a_3\left(16\d q^3+3p^2\right)+\a_4.
\end{eqnarray}
4 - Number of parameters: 10
\begin{eqnarray}
    &L=&\frac{1}{2}\L\left(p\dot{p}^2+q\dot{q}^2\right)
        -\frac{1}{\L}\left\{h_1(p+q)^3\left[\frac{(p+q)^4}{pq}+4(p-q)^2\right]
        +\frac{h_2(p+q)^5}{pq}
        +h_3\left(\frac{p^2}{q}+\frac{q^2}{p}\right)
        +h_4\left(\frac{1}{q}+\frac{1}{p}\right)
        +h_5(p+q)\right\},\nonumber\\
 &\L=&\a_1(p+q)^3\left[\frac{(p+q)^4}{pq}+4(p-q)^2\right]
      +\frac{\a_2(p+q)^5}{pq}
      +\a_3\left(\frac{p^2}{q}+\frac{q^2}{p}\right)
      +\a_4\left(\frac{1}{q}+\frac{1}{p}\right)
      +\a_5(p+q).
\end{eqnarray}
5 - Number of parameters: 11
\begin{eqnarray}
    &L=&\frac{1}{2}\L\left(p^4\dot{p}^2+q^4\dot{q}^2\right)
        -\frac{1}{\L}\left\{\left(\d p^2+q^2\right)
         \left\{h_1\left[4(\d p^2+q^2)^2+\left(\frac{p^3}{q}-\frac{\d q^3}{p}\right)^2\right]
                +h_2\left(\frac{\d q^2}{p^2}+\frac{p^2}{q^2}\right)+h_3\right\}
          +\frac{h_4}{p^2}+\frac{h_5}{q^2}\right\},\nonumber\\
\nonumber\\
 &\L=&\left(\d p^2+q^2\right)
         \left\{\a_1\left[4(\d p^2+q^2)^2+\left(\frac{p^3}{q}-\frac{\d q^3}{p}\right)^2\right]
                +\a_2\left(\frac{\d q^2}{p^2}+\frac{p^2}{q^2}\right)+\a_3\right\}
                -\frac{\a_4}{p^2}-\frac{\a_5}{q^2}.
\end{eqnarray}
6 - Number of parameters: 12
\begin{eqnarray}
&L=&\frac{1}{2}\L\left[\frac{\dot{p}^2}{\sqrt{a_2p^2+a_0}}
                         +\frac{\dot{q}^2}{\sqrt{a_2q^2+a_0}}\right]\nn
    &  &-\frac{\n a_2q^3p+q^2\left[5\n (3a_2p^2+2a_0)+\g a_2p\right]
               +q\left[15\n a_2p^3+6\g a_2p^2+(2\n a_0+h_2)p+4\g a_0\right]
               +\n a_2p^4+\g a_2p^3+h_2p^2+h_1p+h_0}
              {\sqrt{a_2p^2+a_0}}\nn
    &  &-\frac{\n a_2p^3q+p^2\left[5\n (3a_2q^2+2a_0)+\g a_2q\right]
               +p\left[15\n a_2q^3+6\g a_2q^2+(2\n a_0+h_2)q+4\g a_0\right]
               +\n a_2q^4+\g a_2q^3+h_2q^2+h_1q+h_0}
              {\sqrt{a_2q^2+a_0}},\nn
    &\L=& \frac{\a a_2q^3p+q^2\left[5\a (3a_2p^2+2a_0)+\b a_2p\right]
               +q\left[15\a a_2p^3+6\b a_2p^2+(2\a a_0+\a_2)p+4\b a_0\right]
               +\a a_2p^4+\b a_2p^3+\a_2p^2+\a_1p+\a_0}
              {\sqrt{a_2p^2+a_0}}\nn
    &  &+\frac{\a a_2p^3q+p^2\left[5\a (3a_2q^2+2a_0)+\b a_2q\right]
               +p\left[15\a a_2q^3+6\b a_2q^2+(2\a a_0+\a_2)q+4\b a_0\right]
               +\a a_2q^4+\b a_2q^3+\a_2q^2+\a_1q+\a_0}
              {\sqrt{a_2q^2+a_0}}.
\end{eqnarray}
7 - Number of parameters: 13
\begin{eqnarray}
    &L=&\frac{1}{2}\L\left[\frac{\dot{p}^2}{\sqrt{a_2p^2+a_0}}
                      +\frac{\dot{q}^2}{\sqrt{a_2q^2+b_0}}\right]
        -\frac{\n a_2q^2p+q\left[2\n (3a_2p^2+2a_0)+h_2p\right]+\n a_2p^3+h_2p^2+h_1p+h_0}
              {\sqrt{a_2p^2+a_0}}\nn
    &  &-\frac{\n a_2p^2q+p\left[2\n (3a_2q^2+2b_0)+h_2q\right]+\n a_2q^3+h_2q^2+h_1q+k_0}
              {\sqrt{a_2q^2+b_0}},\nn
    &\L=& \frac{\a a_2q^2p+q\left[2\a (3a_2p^2+2a_0)+\a_2p\right]+\a a_2p^3+\a_2p^2+\a_1p+\a_0}
              {\sqrt{a_2p^2+a_0}}\nn
    &  &+\frac{\a a_2p^2q+p\left[2\a (3a_2q^2+2b_0)+\a_2q\right]+\a a_2q^3+\a_2q^2+\a_1q+\b_0}
              {\sqrt{a_2q^2+b_0}}.
\end{eqnarray}
8 - Number of parameters: 13
\begin{eqnarray}
    &L=&\frac{1}{2}\L\left[\frac{\dot{p}^2}{\sqrt{a_2p^2+a_1p+a_0}}
                      +\frac{\dot{q}^2}{\sqrt{b_1q+b_0}}\right]\nn
 &  &-\frac{1}{\L}\left[\frac{\n q(b_1^2q^2+3b_0b_1q+3b_0^2)\left(2a_2p+a_1\right)
             +81\n b_1^3p^3+9b_1^2h_2p^2+h_1p+h_0}{\sqrt{a_2p^2+a_1p+a_0}}
        -\left(b_1q+b_0\right)^{3/2}\left(27\n b_1p+h_2\right)\right],\nn
 &\L=&\frac{\a q(b_1^2q^2+3b_0b_1q+3b_0^2)\left(2a_2p+a_1\right)
           +81\a b_1^3p^3+9b_1^2\a_2p^2+\a_1p+\a_0}{\sqrt{a_2p^2+a_1p+a_0}}
          +\left(b_1q+b_0\right)^{3/2}\left(27\a b_1p+\a_2\right).
\end{eqnarray}
9 - Number of parameters: 13
\begin{eqnarray}
    &L=&\frac{1}{2}\L\left[\frac{\dot{p}^2}{\sqrt{a_2p^2+a_1p+a_0}}+\dot{q}^2\right]
        -\frac{1}{\L}\left[\frac{\m q(q+\g)(2a_2p+a_1)+4Ap^2+h_1p+h_0}{\sqrt{a_2p^2+a_1p+a_0}}
                           +8\m p+Aq(q+\g)+k_0\right],\nn
 &\L=&\frac{\b\m q(2a_2p+a_1)+\a_1p+\a_0}{\sqrt{a_2p^2+a_1p+a_0}}+\b Aq+\b_0.
\label{SYSII.09}
\end{eqnarray}
10 - Number of parameters: 14
\begin{eqnarray}
    &L=&\frac{1}{2}\L\left[\frac{\dot{p}^2}{\sqrt{a_2p^2+a_1p+a_0}}+\dot{q}^2\right]
        -\frac{1}{\L}\left[\frac{\n q(q+\m_1)(2a_2p+a_1)
            +4h_2p^2+h_1p+h_0}{\sqrt{a_2p^2+a_1p+a_0}} 
                            +8\n p+h_2q(q+\m_1)+k_0\right],\nn
 &\L=&\frac{\a q(q+\m_1)(2a_2p+a_1)+4 \a_2p^2+\a_1p+\a_0}{\sqrt{a_2p^2+a_1p+a_0}}
         +8\a p+\a_2 q(q+\m_1)+\b_0.
\end{eqnarray}
11 -  Number of parameters: 14
\begin{eqnarray}
    &L=&\frac{1}{2}\L\left[\frac{\dot{p}^2}{\sqrt{a_2p^2+a_0}}
                       +\frac{\dot{q}^2}{\sqrt{b_2q^2+b_0}}\right]\nn
 &  &-\frac{1}{\L}\left[\frac{\n q^2(3a_2p^2+2a_0)+\g a_2 pq+\n b_2p^4+h_2p^2+h_0}
                                {\sqrt{a_2p^2+a_0}}
                           +\frac{\n p^2(3b_2q^2+2b_0)+\g b_2 qp+\n a_2q^4+h_2q^2+k_0}
                                 {\sqrt{b_2q^2+b_0}}\right],\nn
 &\L=&\frac{\a q^2(3a_2p^2+2a_0)+\b a_2 pq+\a b_2p^4+\a_2p^2+\a_0}{\sqrt{a_2p^2+a_0}}
        +\frac{\a p^2(3b_2q^2+2b_0)+\b b_2 qp+\a a_2q^4+\a_2q^2+\b_0}{\sqrt{b_2q^2+b_0}}.
\label{SYSII.11}
\end{eqnarray}
12 - Number of parameters: 15
\begin{eqnarray}
 &L=&\frac{1}{2}\L\left[\frac{\dot{p}^2}{2\sqrt{a_4p^4+a_2p^2+a_0}}
                            +\frac{\dot{q}^2}{2\sqrt{a_4q^4+b_2q^2+b_0}}\right]
         -\frac{1}{\L}\left[
             \frac{\n q^2(4a_4p^4+3a_2p^2+2a_0)+\g qp(2a_4p^2+a_2)+\n b_2p^4+h_2p^2+h_0}
                  {\sqrt{a_4p^4+a_2p^2+a_0}}\right.\nn
  &  &+\left.\frac{\n p^2(4a_4q^4+3b_2q^2+2b_0)+\g pq(2a_4q^2+b_2)+\n a_2q^4+h_2q^2+k_0}
                    {\sqrt{a_4q^4+b_2q^2+b_0}}\right],\nn
  &\L=& \frac{\a q^2(4a_4p^4+3a_2p^2+2a_0)+\b qp(2a_4p^2+a_2)+\a b_2p^4+\a_2p^2+\a_0}
                {\sqrt{a_4p^4+a_2p^2+a_0}}\nn
   &&       +\frac{\a p^2(4a_4q^4+3b_2q^2+2b_0)+\b pq(2a_4q^2+b_2)+\a a_2q^4+\a_2q^2+\b_0}
                {\sqrt{a_4q^4+b_2q^2+b_0}}
\label{SYSII.12}
\end{eqnarray}
13 - Number of parameters: 16
\begin{eqnarray}
&L=&\frac{1}{2}\Lambda \left[ \frac{\dot{p}^{2}}{\sqrt{%
a_{3}p^{3}+a_{2}p^{2}+a_{1}p+a_{0}}}+\frac{\dot{q}^{2}}{\sqrt{%
b_{2}q^{2}+b_{0}}}\right]   \notag \\
&&-\frac{1}{\Lambda }\left[ \frac{4\nu
q^{2}(3a_{3}p^{2}+2a_{2}p^{2}+a_{1})+64\nu
b_{2}p^{3}+4h_{2}p^{2}+h_{1}p+h_{0}}{\sqrt{a_{3}p^{3}+a_{2}p^{2}+a_{1}p+a_{0}%
}}\right. +\left. \frac{16\nu p\left( 3b_{2}q^{2}+2b_{0}\right) +\nu
a_{3}q^{4}+h_{2}q^{2}+k_{0}}{\sqrt{b_{2}q^{2}+b_{0}}}\right] ,  \notag \\
&\L=&\frac{4\alpha q^{2}(3a_{3}p^{2}+2a_{2}p+a_{1})+64\alpha
b_{2}p^{3}+4\alpha _{2}p^{2}+\alpha _{1}p+\alpha _{0}}{\sqrt{%
a_{3}p^{3}+a_{2}p^{2}+a_{1}p+a_{0}}}+\frac{16\alpha p\left( 3b_{2}q^{2}+2b_{0}\right) +\alpha
a_{3}q^{4}+\alpha _{2}q^{2}+\beta _{0}}{\sqrt{b_{2}q^{2}+b_{0}}}
\label{SYSII.13}
\end{eqnarray}
14 - Number of parameters: 21.
\begin{eqnarray}
&L=&\frac{1}{2}\Lambda \left[ \frac{\dot{p}^{2}}{\sqrt{%
a_{4}p^{4}+a_{3}p^{3}+a_{2}p^{2}+a_{1}p+a_{0}}}+\frac{\dot{q}^{2}}{\sqrt{%
a_{4}q^{4}+b_{3}q^{3}+b_{2}q^{2}+b_{1}q+b_{0}}}\right]  \notag \\
&&-\frac{1}{\Lambda }\left[ \frac{\nu q\left(
4a_{4}p^{3}+3a_{3}p^{2}+2a_{2}p+a_{1}\right) +\nu
b_{3}p^{3}+h_{2}p^{2}+h_{1}p+h_{0}}{\sqrt{%
a_{4}p^{4}+a_{3}p^{3}+a_{2}p^{2}+a_{1}p+a_{0}}}+\frac{\nu p\left( 4a_{4}q^{3}+3b_{3}q^{2}+2b_{2}q+b_{1}\right)
+\nu a_{3}q^{3}+h_{2}q^{2}+k_{1}q+k_{0}}{\sqrt{%
a_{4}q^{4}+b_{3}q^{3}+b_{2}q^{2}+b_{1}q+b_{0}}}\right] ,  \notag \\
&\L=&\frac{\alpha q\left( 4a_{4}p^{3}+3a_{3}p^{2}+2a_{2}p+a_{1}\right)
+\alpha b_{3}p^{3}+\alpha _{2}p^{2}+\alpha _{1}p+\alpha _{0}}{\sqrt{%
a_{4}p^{4}+a_{3}p^{3}+a_{2}p^{2}+a_{1}p+a_{0}}} +\frac{\alpha p\left( 4a_{4}q^{3}+3b_{3}q^{2}+2b_{2}q+b_{1}\right) +\alpha
a_{3}q^{3}+\alpha _{2}q^{2}+\beta _{1}q+\beta _{0}}{\sqrt{%
a_{4}q^{4}+b_{3}q^{3}+b_{2}q^{2}+b_{1}q+b_{0}}}.
\label{SYSII.14}
\end{eqnarray}
}\noindent
\rule[-0.1cm]{20.65cm}{0.01cm}\\[-.45cm]
\rule[-0.1cm]{20.65cm}{0.03cm}
\end{landscape}

\bigskip The system number 14 in Table II is the \emph{master} system
enjoying the maximum number of 21 parameters. It was introduced in 2006 \cite%
{Yehia:2006b}. The system occuring in Table II as number 12 were obtained
recently in \cite{Yehia:2012}. The remaining 12 systems are new.

\end{document}